\documentclass[a4paper,aps,prl,showpacs,twocolumn]{revtex4}%
\usepackage{amssymb}
\usepackage{graphicx}
\usepackage{amsmath}
\usepackage{amsbsy}
\usepackage{amsthm}
\usepackage{bbm}
\usepackage{bm}
\usepackage{epsfig}
\usepackage{mdwlist}
\usepackage{geometry}
\usepackage{amsfonts}%
\providecommand{\U}[1]{\protect\rule{.1in}{.1in}}

\newcommand{\be}{\begin{equation}}
\newcommand{\ee}{\end{equation}}
\newcommand{\beq}{\begin{eqnarray}}
\newcommand{\eeq}{\end{eqnarray}}

\begin{document}
\title{Purification of genuine multipartite entanglement}
\author{Marcus Huber$^{1}$, Martin Plesch$^{2,3}$}
\affiliation{$^{1}$Faculty of Physics, University of
Vienna, Vienna, Austria} \affiliation{$^{2}$Faculty of
Informatics, Masaryk University, Brno, Czech Republic}
\affiliation{$^{3}$Institute of Physics, Slovak Academy of
Sciences, Bratislava, Slovakia}
\date{12 May 2011}

\begin{abstract}
In tasks, where multipartite entanglement plays a central role, state
purification is, due to inevitable noise, a crucial part of the procedure. We
consider a scenario exploiting the multipartite entanglement in a
straightforward multipartite purification algorithm and compare it to
bipartite purification procedures combined with state teleportation. While
complete purification requires an infinite amount of input states in both
cases, we show that for an imperfect output fidelity the multipartite
procedure exhibits a major advantage in terms of input states used.

\end{abstract}

 \maketitle


\section{Introduction }Entanglement is one of the most
striking phenomena of quantum physics. While the primary
interest was focused mainly on the two party case (for an
overview see \textit{e.g.} Ref.~\cite{horodeckiqe}), recent
results show that with a growing system size multipartite
entanglement is of great relevance in contemporary quantum
theory (for an overview see \textit{e.g. }Ref.
\cite{guehnewit}).

Entanglement is inherently manifested in various physical
systems, reaching from condensed matter systems (see
\textit{e.g.} Ref.~\cite{phase}), over ionization of
quantum gases (see \textit{e.g.} Ref.~\cite{helium}) to
small biological systems (see \textit{e.g.}
Refs.~\cite{Caruso,bio}). Also in quantum information
processing (QIP) tasks it is in many cases at the heart of
their advantage over the respective classical counterparts.
It is the crucial resource for measurement based quantum
computing (see \textit{e.g.} Ref.~\cite{qc}), is involved
in most of the popular quantum algorithms (see
\textit{e.g.} Ref.~\cite{brussalgo}) and enables
multi-party quantum communication (see \textit{e. g.}
Refs.~\cite{HBB,Markham}).

Recent advances have made it possible to discriminate
genuinely multipartite entangled states from partially
separable ones (see $\mathit{e.g.}$
Refs.~\cite{horodeckicrit, wocjancrit,
yucrit,hassancrit,hhk}). Some of these criteria are in
large systems implementable even in an experimentally
feasible way and are very robust against noise (see
\textit{e.g.} \cite{seevinckcrit, guehnecrit,HMGH1,GHH1}).
This verification enables secure multipartite quantum
communication protocols such as quantum secret sharing (see
$\mathit{e.g.}$ Refs. \cite{HBB,SHH3}).
This protocol however refers to perfectly noiseless and
error free communication procedures, which even at medium
distances and realistic lab conditions can never be
achieved.
Therefore a purification (also referred to as distillation
in other contexts) procedure was developed, where several
noisy copies of the shared state are employed in an
iterative algorithm to reduce the noise of a subset of
these. If the purification succeeds, eavesdropping and
cheating can be excluded with certainty, thus providing
secure quantum communication even in realistic lab
conditions.

Surprisingly, the purification of genuine multipartite
entanglement does not require the state of the system to
exhibit any genuine multipartite entanglement at all (i.e.
is possible for states which can be decomposed as a convex
sum of biseparable pure states). As long as one party is
able to purify bipartite entanglement with respect to all
other parties, it is possible two create any kind of
genuinely multipartite entangled states by means of quantum
teleportation (as pointed out in Ref.~\cite{guehnewit}).
Multipartite purification procedures of states affected by
bipartite noisy channels have been intensively investigated
in Refs.~\cite{dur1,dur2,dur3,dur4,dur5,dur6,dur7} and
general distillation of
multipartite entanglement has been studied in Refs.~\cite{Murao,ChenLo,smolin}%
.

In this paper we consider a generalized purification
algorithm for multipartite states, which is based on the
procedure originally introduced in Ref. \cite{Bennet} and
is similar to the algorithm introduced in
Ref.~\cite{Augusiak}. In our setting parties share a
multipartite (imperfectly) entangled state, which can be a
result of a previously obtained state, which has undergone
decoherence, or a state, which was transferred via a noisy
multipartite channel. We compare the efficiency of this
protocol with its bipartite counterpart in the means of the
number of resource states used to get a single, highly
entangled output state.

Let us begin by concisely defining the setup: N parties (among them Dora the
dealer) wish to implement a QIP task employing genuine multipartite
entanglement. They have prepared and shared a genuinely multipartite entangled
state $|GHZ\rangle=\frac{1}{\sqrt{2}}\left(  |0\rangle^{\otimes N}%
+|1\rangle^{\otimes N}\right)  \,$, also known as GHZ
(Greenberger-Horne-Zeilinger) state (this specific state is \textit{e.g.} used
in the quantum secret sharing protocols in Refs.~\cite{HBB,SHH3}). However,
due to decoherence or noise in the multipartite channel used for distributing
the qubits the parties end up with a noisy mixed state
\begin{equation}
\rho_{in}=q\left\vert GHZ\right\rangle \left\langle GHZ\right\vert +\frac
{1-q}{2^{N}}\openone. \label{rho_in}%
\end{equation}
The fidelity of this state with a pure GHZ state is
$F_{in}=q+\frac{\left( 1-q\right)  }{2^{N}}=\frac{\left(
2^{N}-1\right)  q+1}{2^{N}}$ and in-fidelity
$\delta=1-F_{in}=\frac{\left(  2^{N}-1\right)
}{2^{N}}\left( 1-q\right)  $. This fidelity might not be
sufficient for performing the QIP task (e.g. secure quantum
secret sharing) reliably. So the parties have to use more
copies of the state (\ref{rho_in}) and perform purification
on them to obtain a single, high quality, state. We will
express the effectiveness of the protocol as the number of
source multipartite states (\ref{rho_in}) needed to obtain
a single highly entangled state with a fidelity at least
$1-\varepsilon$.

\section{Bipartite procedure }We proceed as follows:
Dora prepares bipartite entangled states with all other
parties. Secondly, Dora performs bipartite distillation
protocols with all other parties to obtain pure enough
states for stage three. In this stage, Dora locally
prepares a GHZ state of $N$ qubits and use the bipartite
entangled states to teleport all the qubits of this state
but one to all other parties.

\subsection{Preparation of the bipartite state }Let us
assume that Dora wants to prepare a bipartite entangled
state with one of the other parties, Julia. She selects one
of the source states and asks all other parties to perform
a measurement in the $\left\vert \pm\right\rangle $ basis.
Depending on the outcomes of the measurements of the other
parties, the bipartite state shared between Dora and Julia
will be $\rho_{DJ}=q\left\vert \Phi^{\pm}\right\rangle
\left\langle \Phi^{\pm}\right\vert +\frac{1-q}{4}\openone$,
where $\left\vert \Phi^{\pm}\right\rangle
=\frac{1}{\sqrt{2}}\left( \left\vert 00\right\rangle
\pm\left\vert 11\right\rangle \right)  $ are Bell states
and the sign is given by the parity of outcomes of the
state $\left\vert -\right\rangle $ of the measurement. Dora
collects all the information about the measurement outcomes
from all parties except Julia and in the case of an odd
number of the $\left\vert -\right\rangle $ results she will
perform a local phase operation to correct the sign. In
such a way Dora prepares $M$ bipartite states with each of
the $N-1$ remaining parties, requiring $M\left( N-1\right)
$ source states.

\subsection{Bipartite distillation }As a second step, Dora
performs distillation of entanglement on all $M$ bipartite
entangled pairs shared with Julia. Using the procedure
described in \cite{Bennet}, in every step they can
concentrate entanglement from two pairs to one pair by
performing a C-NOT operation on pairs of qubits on both
sides and measuring the target qubits. After this
operation, the parameter $q$ changes to
\begin{equation}
q\rightarrow\frac{4q^{2}+2q}{3\left(  q^{2}+1\right)  }. \label{distillation}%
\end{equation}

Dora and Julia have used two pairs of bipartite source
states between them and the result $0$ is obtained with
probability $\frac{1}{2}$ on Doras' side. The probability
of measuring $0$ in Julias' side depends on the parameter
$q$ and is $\frac{1+q}{2}$. Thus, for every step they need
in average $\frac{8}{1+q}$ source states, which can be
bounded from below by $4$. With $k$ steps for all parties,
they have used more than $4^{k}\left( N-1\right) $ states
to obtain $N-1$ pairs of highly entangled states with the
parameter $q_{k}$ given by the recursive application of the
formula (\ref{distillation}).

\subsection{Reexportation }We use the distilled states
to perform reexportation \cite{Teleportation} of individual
qubits of a GHZ state locally prepared by Dora. As the
shared entanglement is not perfect, neither will the
reexportation be. Dora performs a Bell measurement on a
pair of qubits, where one qubit will be taken from the GHZ
state and the other from the bipartite state shared with
Julia. She communicates the result consisting of two bits
to Julia and Julia performs corrective transformation, if
needed. After repeating this procedure with all other
parties, the shared state will have the form
\begin{align*}
\rho_{f}  &  =q_{k}^{N-1}\left\vert GHZ\right\rangle \left\langle
GHZ\right\vert +\\
&  \sum_{j=1}^{N-1}\left(  1-q_{k}\right)  ^{j}q_{k}^{N-j-1}\sum_{i=1}%
^{\binom{N-1}{j}}2^{-j}\openone_{B(j,i)}\otimes\rho_{rest}^{diag}\\
\rho_{s}^{diag}  &  =\frac{1}{2}(|0\rangle\langle0|^{\otimes s}+|1\rangle
\langle1|^{\otimes s}),
\end{align*}
where $B(j,i)$ is a set of $j$ qubits, selected by the choice of $i$ and we
sum over all possible sets of qubits. The fidelity of this state with the GHZ
state is
\begin{align}
F_{out}  &  =q_{k}^{N-1}+\sum_{j=1}^{N-1}\left(  1-q_{k}\right)  ^{j}%
q_{k}^{N-j-1}\binom{N-1}{j}2^{-j-1}\nonumber\\
&  =\frac{1}{2}\left[  q_{k}^{N-1}+\left(  \frac{1+q_{k}}{2}\right)
^{N-1}\right]  \label{outout}%
\end{align}
and is expected to be at least $1-\varepsilon$.

\subsection{Small errors }We will analyze the results for
the case when $\delta$ is small enough (hence $q$ close to
$1$). Substituing $\Delta=1-q$, the transformation formula
for distillation (\ref{distillation}) changes to
\begin{align}
\Delta\rightarrow\frac{4\Delta-4\Delta^{2}}{6-6\Delta+3\Delta^{2}},
\end{align}
which simplifies to $\Delta\rightarrow\frac{2}{3}\Delta$.
Substituing back we get $q\rightarrow\frac{1}{3}\left(
1+2q\right)  $\emph{ }and we can calculate the recurrence
directly obtaining (after $k$ distillation steps)
$q_{k}=1-\left( \frac{2}{3}\right)  ^{k}\left(  1-q\right)
$. Substituting to the approximated version of
(\ref{outout}) we get
\[
\left(  N-1\right)  \left(  1-q\right)  \left(  \frac{2}{3}\right)  ^{k-1}%
\leq\varepsilon
\]
and%
\[
k\geq1+\log\left[  \frac{2^{N}}{2^{N}-1}\frac{\delta}{\varepsilon
}(N-1)\right]  /\log\left[  \frac{3}{2}\right]
\]
with logarithm taken in basis 2. The overall number of
states needed will thus be at least%
\[
4\left(  N-1\right)  \left[  \frac{2^{N}}{2^{N}-1}\frac{\delta}{\varepsilon
}(N-1)\right]  ^{\frac{2}{\log\left[  3/2\right]  }},
\]
which can be approximated for large $N$ as%
\[
4N^{4.42}\left(  \frac{\delta}{\varepsilon}\right)  ^{3.42}.
\]

\subsection{Large errors }For larger errors the
approximations do not work and it is not possible to
calculate an explicit expression for the recurrence formula
(\ref{distillation}). However, numerical analysis for
different values of $\delta$ and $N$ for a fixed
$\varepsilon=0,01$ show a strong dependence of the number
of source states needed on $N$.

\subsection{Smart distillation }More advanced distillation
protocols mentioned in Ref.~\cite{Bennet} suggest that it
might be possible to perform more effective protocols, if
the fidelity of the input state with the expected entangled
state is known. In this case, the fraction of states lost
in every distillation step depends on the fidelity of the
source states and might be rather low. But even if we had
allowed Dora and Julia to use a virtually ideal distillation
protocol, allowing them to extract all bipartite
entanglement from the shared state (given by the
entanglement of formation), the bipartite procedure would
still have been highly inefficient due to large loss on
fidelity given by (\ref{outout}).

\section{Multipartite procedure } In the multipartite case,
all participants employ local operations and classical
communication (LOCC) to purify $\rho_{in}$ directly.
The generalization of the bipartite distillation procedure
introduced in Ref.~\cite{Bennet} works as follows:
\begin{itemize}
\item All parties take two copies of the input state \item
They label the first qubit target and the second control
and perform a CNOT-gate operation on their two qubits \item
They measure their target qubit in $\sigma_{z}$ basis, keep
their control qubit if the outcome is $0$, and dismiss it
otherwise \item Finally they communicate their results, and
only keep those states, where they all shared the outcome
$0$ and dismiss all others. In this case the state exhibits
stronger multipartite entanglement than the source
state.\vspace{-3mm}
\end{itemize}
Except for the fact that multipartite distillation directly uses
multipartite entanglement provided in the source,
without converting it to bipartite entanglement and back,
it can also by performed in a much simpler way, compared to
the bipartite one Ref.~\cite{Bennet}. Randomized rotations
between every iteration step are not necessary, if at least
three-partite entanglement is distilled.

\subsection{Distillation operation} We define the map
\[
\Lambda\lbrack\rho^{\otimes2}]:=\text{Tr}_{c_{1}c_{2}\cdots c_{n}}%
[D\rho^{\otimes2}D^{\dagger}]
\]
with
\[
D:=(\mathbbm{1}\otimes|0\rangle\langle0|)^{\otimes n}CNOT^{\otimes n},
\]
where the partial trace is taken over all target qubits. As
\begin{align*}
\Lambda\lbrack\mathbbm{1}^{\otimes2}]  &  =\mathbbm{1}\\
\Lambda\lbrack|GHZ\rangle\langle GHZ|^{\otimes2}]  &  =\frac{1}{2}%
|GHZ\rangle\langle GHZ|\\
\Lambda\lbrack|GHZ\rangle\langle GHZ|\otimes\mathbbm{1}]  &  =\Lambda
\lbrack\mathbbm{1}\otimes|GHZ\rangle\langle GHZ|]=\rho_{n}^{diag},
\end{align*}
the initial state
\[
\rho_{in}=q\left\vert GHZ\right\rangle \left\langle GHZ\right\vert +\frac
{1-q}{2^{N}}\mathbbm{1}
\]
transforms to
\[
\Lambda\lbrack\rho^{\otimes2}]=\frac{q^{2}}{2}\left\vert GHZ\right\rangle
\left\langle GHZ\right\vert +\frac{(1-q)^{2}}{2^{2N}}\mathbbm{1}+\frac
{2q(1-q)}{2^{N}}\rho_{n}^{diag}.
\]
The success probability of this procedure (probability of
obtaining the result $0$ from measurements of all parties)
depends on the initial state and is given by:
\begin{equation}
P_{succ}=\text{Tr}\left[  \Lambda\left[  \rho^{\otimes2}\right]  \right]
=\frac{q^{2}}{2}+\frac{2q(1-q)}{2^{N}}+\frac{(1-q)^{2}}{2^{N}}%
\end{equation}
For large $N$ it is always sufficient to apply the protocol
once and the result is already arbitrarily close to the GHZ
state. For intermediate $N$, however, it might be necessary
to repeat the procedure more times. In such a case the
input state will be of a different form
\begin{equation}
\rho_{i}=q\left\vert GHZ\right\rangle \left\langle GHZ\right\vert +\frac
{r}{2^{N}}\mathbbm{1}+s\rho_{n}^{diag}. \label{rho_i}%
\end{equation}
Using the additional relations
\begin{align*}
\Lambda\lbrack\mathbbm{1}\otimes\rho_{n}^{diag}]  &  =\Lambda\lbrack\rho
_{n}^{diag}\otimes\mathbbm{1}]=\rho_{n}^{diag}\\
\Lambda\lbrack|GHZ\rangle\langle GHZ|\otimes\rho_{n}^{diag}]  &
=\Lambda\lbrack\rho_{n}^{diag}\otimes|GHZ\rangle\langle GHZ|]=\frac{1}{2}%
\rho_{n}^{diag}\\
\Lambda\lbrack\rho_{n}^{diag}\otimes\rho_{n}^{diag}]  &  =\frac{1}{2}\rho
_{n}^{diag}%
\end{align*}
we can continue the iteration using recurrence relations
\begin{equation}
q\rightarrow\frac{q^{2}}{2},\; r\rightarrow\frac{r^{2}}{2^{N}},\;
s\rightarrow\frac{s^{2}+sq}{2}+\frac{sr+2qr}{2^{N}}%
\end{equation}
until we reach the desired fidelity with the $GHZ$-state.
This result implies that the $GHZ$-state remains
distillable up to a noise threshold of
$q_{noise}=\frac{2}{2^{n}-2}$. This is remarkable as it was
recently shown using semidefinite programming that e.g. the
four party $GHZ$-state is biseparable up to a noise
threshold of $q_{noise}\approx0.46$ in
Ref.~\cite{Taming}.

\begin{figure}[ptb]
\begin{center}
\includegraphics[
width=3in ]{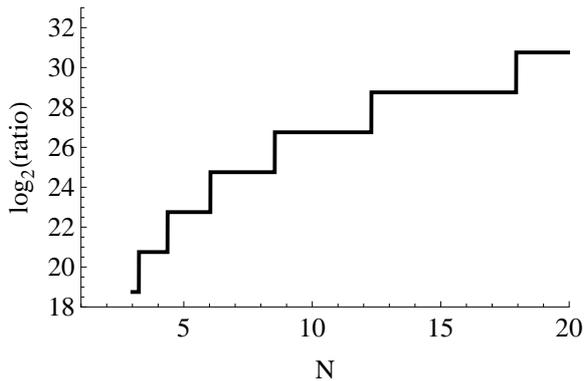}
\end{center}
\caption{Logarithm (base 2) of the ratio between the number of source states
with in-fidelity $\delta=0.2$ needed to get a single GHZ state with
in-fidelity $\varepsilon=0.01$ for different values of $N$, using bipartite
and multipartite distillation. The ratio even for moderate values of $N$
exceeds $10^{6}$. }%
\label{figure1}%
\end{figure}

\begin{figure}[ptb]
\begin{center}
\includegraphics[
width=3in ]{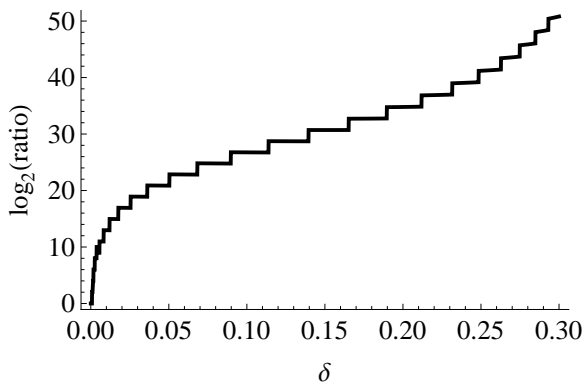}
\end{center}
\caption{Logarithm (base 2) of the ratio between the number of source states
needed to get a single GHZ state with in-fidelity $\varepsilon=0.01$ for $10$
parties and different values of $\delta$, using bipartite and multipartite
distillation.}%
\label{figure2}%
\end{figure}

The number of resource states needed for the whole
procedure is then given by the inverse of the norm of the
state (\ref{rho_i}) $M=\frac{1}{(q+r+s)}$. In Figures
\ref{figure1} and \ref{figure2} we show the ratio of the
number of resource states needed to reach a given output
fidelity of a single state. It can be clearly seen that the
multipartite procedure is far more effective even for a
moderate number of parties $N$ and for the whole scale of
input fidelities. What is especially important, the
distillation protocol produces high-fidelity GHZ states
after a single, or a few steps even in the case of very low
input fidelities (e.g. $\delta=0.5$), where the bipartite
distillation protocol is completely useless.

\section{Conclusions} In this paper we have shown an example of a
scenario, in which direct multipartite entanglement
purification is far more efficient than bipartite
distillation protocols combined with state teleportation.
For a moderate number of parties the multipartite protocol
gives reasonable results even with extremely high input
noise (e.g. reaching up to $\approx99.8\%$ for $n=10$). The
amount of resources needed using the bipartite protocol
grows much faster with the input noise than in the
multipartite case. The nature of the advantage stems mainly
from the structure of the noise, which has inherently
multipartite features.

Furthermore, this paper provides evidence of biseparable
states being distillable to genuinely multipartite
entangled ones without resorting to state teleportation.

In a different context, using bipartite noisy communication
channels, there has been intensive study of a very similar
scenario in Refs.~\cite{dur1,dur4,dur5,dur7}, also
comparing the effectiveness of the two different
strategies. In this case there is a noise regime where the
multipartite purification approach still outperforms the
bipartite one, however, the advantage never reaches the
level obtained for multipartite channels. This leads to a
surprising conclusion that multipartite entangled states
affected by noise in multipartite channels are far easier
and more effectively purifiable than the same states
affected by local noise.

\section{Acknowledgments} We thank Ch. Spengler and A.
Gabriel for inspiring discussions. This work was supported
by the SoMoPro project SIGA 862 and by the CE SAS QUTE. The
collaboration is a part of \"{O}AD/APVV SK-AT-0015-10
project. M. Huber gratefully acknowledges the support of
the Austrian Fund project FWF-P21947N16.

\end{document}